# Evaluation of Doppler Shifts to Improve the Accuracy of Primary Atomic Fountain Clocks

Jocelyne Guéna,[1] Ruoxin Li,[2] Kurt Gibble,[1,2] Sébastien Bize,[1] and André Clairon[1]
[1]LNE-SYRTE, Observatoire de Paris, CNRS, UMPC, 61 avenue de l'Observatoire, 75014 Paris, France
[2]Department of Physics, The Pennsylvania State University, University Park, PA 16802, USA

We demonstrate agreement between measurements and *ab initio* calculations of the frequency shifts caused by distributed cavity phase variations in the microwave cavity of a primary atomic fountain clock. Experimental verification of the finite element models of the cavities gives the first quantitative evaluation of this leading uncertainty and allows it to be reduced to $\delta\nu/\nu = \pm 8.4 \times 10^{-17}$. Applying these experimental techniques to clocks with improved microwave cavities will yield negligible distributed cavity phase uncertainties, less than $\pm 1 \times 10^{-17}$.

Atomic clocks deliver the most accurate measurements of any physical observable, frequency and time. Their precision enables global positioning systems and stringent tests of fundamental physics. The most accurate clocks realize the definition of the SI second and the widely-used international atomic time, TAI. The accuracy of TAI comes from an ensemble of laser-cooled atomic fountain clocks from around the world, which are currently limited by first order Doppler shifts [1-7]. These Doppler shifts occur when the cold but moving atoms interact with a field inside the microwave cavity that has a spatial phase variation (Fig. 1a) because it is not a totally pure standing-wave. The scale for Doppler shifts is extremely large, 1 part in $10^8$. The shift is highly suppressed by a velocity reversal from gravity in Fig. 1a, and the purity of the standing wave. Uncertainty estimates are as low as $\pm 3 \times 10^{-16}$, excepting a few cases where it is incorrectly evaluated [1-7]. Here, we present precise measurements of a primary clock's frequency and large finite element calculations of the cavity fields. With no free parameters, these measurements are the first stringent test and quantitative confirmation of the key behaviors of this currently dominant systematic error. The validation of the model allows us to significantly reduce our Doppler uncertainty, to $8.4 \times 10^{-17}$, and shows that optimized cavities will have negligible phase variations yielding uncertainties less than $\pm 1 \times 10^{-17}$, thus removing the largest barrier to significantly better primary clock accuracy.

This Doppler shift is known as the distributed cavity phase (DCP) frequency shift. Although DCP shifts have been considered for 35 years, no measurements have reasonably agreed with calculations [8]. There has been little progress because the phase of the intra-cavity field cannot be accurately mapped and the calculations are difficult since the holes in the cavities, required for the atomic traversals, produce large perturbations [9]. Frequency measurements have given indirect information about the phase variations, but often the associated frequency shifts were misattributed to other systematic errors [5]. Recently, a combination of finite element and analytic models have elucidated the behaviors of the fields and their DCP shifts [7,9].

Large, densely-meshed finite element calculations are required to accurately calculate the fields in clock cavities. All fountains clocks contributing to TAI use cylindrical $TE_{011}$, centimeter-size cavities with feeds at the cavity midplane and holes in the endcaps (Fig.1a). The sharp "corner" of these holes produce nearly singular fields, to a scale as small as the skin depth [9], 0.7 μm for copper at the cesium clock transition frequency, 9.2 GHz. This leads to the requirement for dense meshing, which makes direct 3D solutions unfeasible. However, the cylindrical symmetry allows the field to be expressed as an azimuthal Fourier series of 2D finite element solutions [7,9]. We write the microwave magnetic field as the sum of a large standing wave $\mathbf{H}_0(\mathbf{r})$ and a small field, $\mathbf{g}(\mathbf{r})=\Sigma g_m(\rho,z) \cos(m\phi)$, which describes the cavity feeds and wall losses [7].

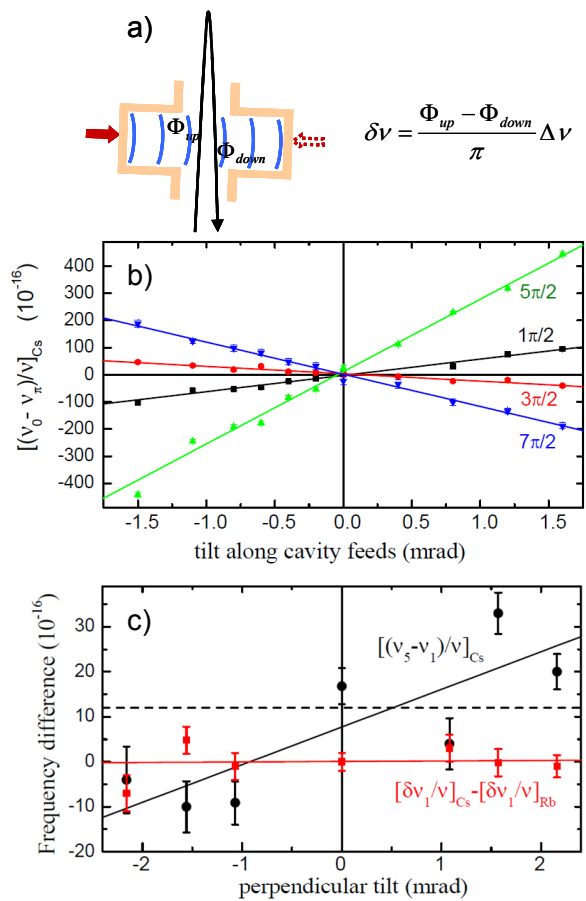

**Fig. 1.** a) Schematic of distributed cavity phase (DCP) frequency shifts. Feeding a microwave cavity from only one side produces a phase gradient in the cavity and gives a Doppler shift if the atomic fountain is tilted. b) Frequency differences for feeding at $\phi=0$ or $\pi$ versus tilt along the cavity feeds for 1,3,5, and $7\pi/2$ pulses. All differences are 0 when the fountain has no tilt. c) The black circles are the frequency differences for $5\pi/2$ pulses and $\pi/2$ pulses versus tilt perpendicular to the feeds, suggesting an inhomogeneous surface resistance. The dashed line is the predicted $m=0$ DCP shift. The red squares are the differences of the Cs and Rb clock frequencies for $\pi/2$ pulses, consistent with no $m=1$ DCP shift for perpendicular tilts.

$$\mathbf{H}(\mathbf{r}) = \mathbf{H}_0(\mathbf{r}) + \left(\frac{2\Delta\omega}{\Gamma} + i\right)\mathbf{g}(\mathbf{r}) \quad (1)$$

Here $\Delta\omega$ is the detuning of the cavity from resonance and $\Gamma$ is the cavity fullwidth. The fields have an $e^{-i\omega t}$ time dependence and the phase of the field is $\Phi \approx -g_z(\mathbf{r})/H_{0z}(\mathbf{r})$, since only the $\hat{z}$ component couples to the atoms [6,7]. The $g_m(\rho,z)$ components are proportional to $\rho^m$ for $\rho \to 0$ and only 3 terms of this series, m=0, 1, and 2, contribute significantly because the atoms pass through the cavity near its axis. These terms describe power flow from the feed(s) to the walls with various symmetries: m=0, to the endcap walls, m=1, from one side of the cavity to the other, and m=2, a quadrupolar flow from opposing feeds. We next describe our experimental techniques to measure and minimize the DCP shift of each azimuthal component in the FO2 dual Cs/Rb fountain clock at SYRTE (Systèmes de Référence Temps-Espace) [1].

*m=0 phase variations:* The $g_0(\rho,z)$ term describes power supplied at the cavity midplane and absorbed on the endcaps. It creates large longitudinal phase gradients, which are azimuthally symmetric for homogeneous endcap surface resistances, and produces frequency shifts that depend strongly on the amplitude b of the microwave field, as shown in the top inset of Fig. 2a. Here, for normal clock operation, power is supplied to both feeds and π/2 pulses on the upward and downward cavity traversals produce maximum Ramsey contrast near b=1 [10]. The frequency shift is singular near b=4, 8, and 10 where the Ramsey fringe contrast goes to 0 since the pulse areas are multiples of π. We therefore plot in Fig. 2 the well-behaved difference in transition probability $\delta P = (\delta P^+ - \delta P^-)/2$, at detunings of $\pm\Delta\nu/2$, where $\Delta\nu$=0.822 Hz is the transition fullwidth. We extract δP from the measured frequency shift and Ramsey fringe contrast. For optimal amplitude, π/2 pulses, the contrast is essentially 100% and $\delta P = \pi \delta\nu/(2\Delta\nu)$.

The DCP shifts in Fig. 2a are small at b=2,6 and large at b=4,8 because the phase is symmetric about the cavity midplane [7]. For 9π/2 pulses, we observe a DCP shift of $6.5 \times 10^{-15}$, 20 times larger than the clock's uncertainty for normal operation. Although this m=0 DCP shift must exist in all current primary clocks, there is only one other report of a comparable shift [11]. These large shifts occur even though there are negligible transverse variations of the phase. Instead, the transverse variation comes from $H_{0z}(\mathbf{r})$, which gives different pulse areas to the expanding cloud on the upward and downward passages [7]. With no free parameters, our measurements versus microwave amplitude quantitatively agree with the calculated atomic response to the finite-element fields [12]. The left inset in Fig. 2a shows that the DCP shift is extremely small at low amplitude, b≤1, because the effective phase is simply the longitudinal average, which has a very small transverse variation [7]. The predicted shift at optimal amplitude is $\delta P=7\times10^{-8}$, corresponding to $\delta\nu/\nu=4\times10^{-18}$, which we take as our m=0 DCP uncertainty.

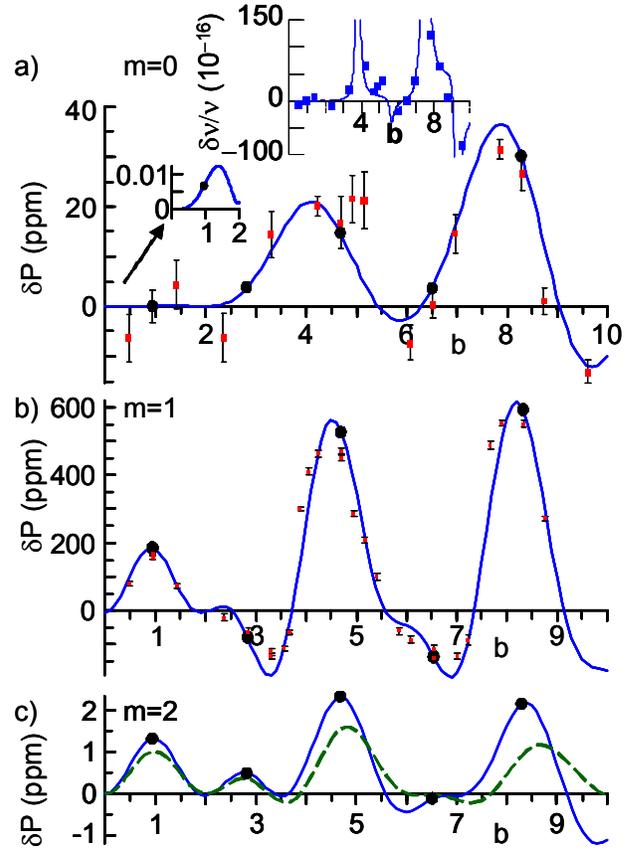

**Fig. 2.** Measured (squares) and calculated (lines) change in transition probability δP from DCP shifts versus microwave amplitude b, where 1,3,5 … π/2 pulses (black dots) are near b=1,3,5 …[10]. a) Azimuthally symmetric m=0 DCP shifts produce a negligible shift at b=1 (left inset) and have a large amplitude dependence. In the top inset, the DCP frequency shift is large and singular near b=4 and 8 (2π and 4π pulses). b) m=1 DCP shifts, given by the clock's frequency difference between feeding the cavity at φ=0 and π with a fountain tilt of 1.6 mrad. Balancing the feeds and nulling the tilt minimizes this shift. c) Predicted m=2 DCP shift for FO2 with an effective 9.9 mm detection laser beam waist (solid) and an initial cloud offset of 2mm at launch with uniform detection (dashed).

To calculate $\delta P$ and the frequency shifts, we independently developed two Monte Carlo simulations. One integrates the atomic response to the fields over each random fountain trajectory while the other first calculates an effective phase $\delta\Phi_m$ for a cavity traversal [7], and then averages $\delta\Phi_m$ over random trajectories. Both calculations include the spatially inhomogeneous detection of the atoms and the apertures in the fountain that cut the atomic clouds. The two simulations agree and, using the measured cloud temperature and size, reproduce the microwave amplitude dependence of the Ramsey fringe contrast and Rabi flopping for the two cavity passages.

*m=1 phase gradients:* The m=1 field component $\mathbf{g}_1(\rho,z)$ produces a phase gradient as depicted in Fig. 1a. If an m=1 component exists, a tilt of the fountain, or an off-center initial launch position, produces a DCP shift because the average positions of the atoms, and the phases of the field, on the two passages, are different. We probe m=1 DCP shifts by intentionally tilting the entire fountain and measuring the frequency difference $\nu_0-\nu_\pi$, by feeding the cavity alternately at $\phi=0$ and $\pi$ [13]. For tilts as large as $\pm1.6$ mrad and (1,3,5,7) $\pi/2$ pulses, the frequency difference in Fig. 1b is proportional to the tilt and vanishes at the same tilt to within $\pm40$ μrad. Using the zero crossing, we can align the fountain more accurately than our early mechanical method and also properly account for any irregularities of the atomic distribution. However, long-term fluctuations of the position and velocity of the atomic cloud limit the alignment to $\pm0.1$ mrad. Fig. 2b shows the m=1 DCP shift, $\nu_0-\nu_\pi$, versus microwave amplitude for a tilt of 1.6 mrad. The magnitude of m=1 DCP shifts is inversely proportional to the loaded cavity Q [7] and very large for a single feed.

During normal clock operation, it is crucial to minimize the potentially large m=1 DCP shifts. We balance the feeds by adjusting their amplitudes so that the clock's frequency, measured against another cesium clock, has no tilt dependence for $\pi/2$ pulses to $\pm2.2\times10^{-16}$ mrad$^{-1}$. We then align the fountain to have no tilt to within $\pm0.1$ mrad (crossing point in Fig. 1b) along the feed axis, reducing this m=1 DCP error to $\pm2.2\times10^{-17}$, limited by our balancing and trajectory fluctuations. Our method differs from the currently widely used technique which balances the amplitude of $H_{0z}(\mathbf{r})$ (pulse area) from each feed but does not eliminate the m=1 tilt sensitivity when the feeds have different reflectivities from their external circuitry. In the past, when we balanced $H_{0z}(\mathbf{r})$, we had a significant residual tilt sensitivity of $8\times10^{-16}$ mrad$^{-1}$.

We also carefully balance the phase of both cavity feeds [6]. While it's accepted that a phase imbalance directly excites $\mathbf{g}_1(\rho,z)$, there are disagreements about the sensitivity [14]. Our model clarifies the disagreements, showing that the DCP shift due to phase imbalances depends on the microwave cavity detuning and vanishes at resonance [7]. In Fig. 3, we experimentally observe this dependence with a large tilt and $\pi/2$ pulses. As expected, there is no shift for a nearly resonant cavity (black) since the $\mathbf{g}_1(\rho,z)$ that is excited in Eq. 1 is imaginary (phase of $\pi/2$) so that $\Phi=\text{Im}[H_z(\mathbf{r})]/\text{Re}[H_z(\mathbf{r})]=0$ [7]. But, as the cavity is detuned (blue and red), each feed acquires an additional phase shift relative to $\mathbf{H}_0(\mathbf{r})$ and the feed with the smaller phase shift supplies more power, yielding $\text{Im}[H_z(\mathbf{r})]=i\Delta\omega/\Gamma\ \mathbf{g}_1(\rho,z)\cos(\phi)$. In Fig. 3 we measure the clock's frequency with a phase difference, $\nu(\Delta\psi)$, and for only a left (right) feed, $\nu_{0(\pi)}$, and plot $\nu(\Delta\psi)-(\nu_0+\nu_\pi)/2$, normalized by $\nu_0-\nu_\pi$. The lines are fits to the prediction $\Delta\omega/\Gamma\ \tan(\Delta\psi/2)$ [7,15,16].

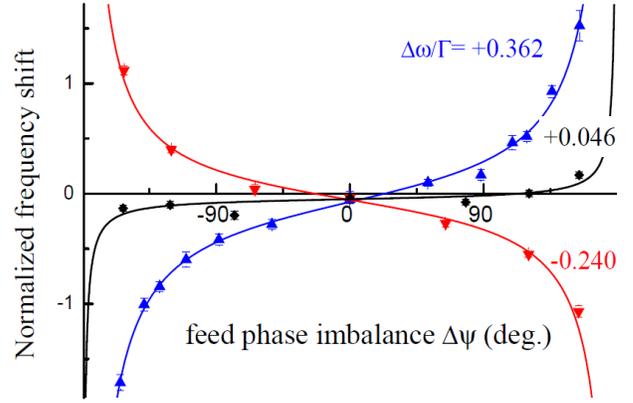

**Fig. 3.** Normalized frequency shift versus phase imbalance of the cavity feeds for a 1.6 mrad tilt along the feeds. As the cavity is tuned to resonance (black), the DCP shift goes to 0. The solid lines are fits to A $\tan(\Delta\psi/2)$ and A agrees with the prediction, $\Delta\omega/\Gamma$.

Without independent feeds at $\phi=\pm\pi/2$, current clocks cannot null the tilt perpendicular to the feeds as above. A DCP shift for perpendicular tilts can only come from inhomogeneous wall losses that produce a $\mathbf{g}_1(\rho,z)\sin(\phi)$. Despite careful machining and special attention to surface finish, Fig. 1c (solid black line) shows a differential perpendicular tilt sensitivity for two $5\pi/2$ pulses relative to $\pi/2$ pulses of $8.4(2.6)\times10^{-16}$ mrad$^{-1}$, suggesting conductivity inhomogeneities in our cavity. This dictates that the perpendicular m=1 DCP shift must be evaluated and it in fact is the largest contribution to our DCP uncertainty. To date, all primary clock evaluations have taken m=1 DCP shifts for perpendicular tilts to be 0 by symmetry [1-6,13]. Finite element modeling suggests that local surface resistances must be at least 20% larger than for a pristine copper surface. The losses at each wall position can produce very different $\delta P(b)$; for example, b=1 and 5 can have the same $\delta P$ or there can be a large $\delta P$ at

b=1 but δP=0 at b=5 [7]. Thus, differential measurements of the clock's frequency versus amplitude and perpendicular tilt alone cannot establish a stringent m=1 DCP uncertainty [17].

Here, we establish a DCP uncertainty for perpendicular tilts by measuring the frequency difference of the FO2 Cs and Rb clocks at optimal amplitude versus a common tilt. The difference in Fig. 1c (red) is $1(8) \times 10^{-17}$ mrad$^{-1}$, consistent with no tilt sensitivity. Then, to null the tilt, we maximize the number of atoms returning through the cavity. An image shows that the initial Cs cloud position is centered to ±2 mm, ensuring no perpendicular tilt within ±0.7 mrad and giving an uncertainty of $\pm 6 \times 10^{-17}$ [18]. A cavity with 4 independent feeds, at $\phi=0, \pi$, and $\pm\pi/2$, would allow a more precise alignment and reduce our m=1 DCP uncertainty [7].

*m=2 phase variations:* While the m=2 quadrupolar phase variation is maximally excited by feeds at $\phi=0$ and $\pi$, its $\rho^2 \cos(2\phi)$ radial dependence gives small phase shifts. If the cloud is small but not centered, the atoms can experience a non-zero phase on the upward cavity traversal relative to the average phase on the downward passage, resulting in a DCP shift (Fig. 2c dashed). On the downward passage, the density is nearly uniform so the average m=2 phase for the entire cloud is zero. However, a spatially inhomogeneous detection effectively modulates this uniform distribution, also producing a DCP shift. We measure the transition probability by imaging the fluorescence from a Gaussian laser beam, giving a higher detection efficiency at $\phi=\pm\pi/2$ than $\phi=0$ and $\pi$. The model (Fig. 2c solid) predicts a DCP clock correction of $-7.5 \times 10^{-17}$. We take half the correction as the uncertainty and add it in quadrature to the uncertainty for a 2mm cloud offset, to get $\pm 5.5 \times 10^{-17}$. This can be made negligibly small by making the imaging uniform, rotating our 2-feed cavity by $\pi/4$, or, preferably, using 4 cavity feeds.

*Conclusions:* With no free parameters we demonstrate agreement between measurements and calculations of the distributed cavity phase (DCP) shift, a first order Doppler shift, in a primary atomic clock. The verification of the model allows a quantitative evaluation and reduction of this currently largest systematic error for the best atomic clocks that define TAI. Three azimuthal components, m=0, 1, 2, produce significant DCP shifts. By evaluating each component, we improve our DCP uncertainty to $\delta\nu/\nu = \pm 8.4 \times 10^{-17}$, limited by the tilt sensitivity to m=1 phase gradients perpendicular to the cavity feeds and m=2 phase variations [19]. We demonstrate the importance of balancing feeds by measuring DCP shifts versus tilt and probing DCP shifts due to inhomogeneous surface resistances. Significant reductions of the DCP uncertainty, to less than $\pm 1 \times 10^{-17}$, will be possible using a cavity with four independent, azimuthally-distributed feeds [7]. This would allow precise alignment of the fountain tilt in both directions to reduce the m=1 DCP shifts, and also make m=2 shifts negligible. Improved cavity designs based on this validated model can further eliminate the m=0 longitudinal phase gradients, even at high microwave amplitudes [7]. Amplitude dependence can then be used to more precisely evaluate several systematic effects, including the atom-interferometric lensing of the atomic wave packets by the cavity's microwave field [20].

We acknowledge the assistance of SYRTE technical services and support from the NSF, Penn State, and la Ville de Paris (KG). SYRTE is UMR CNRS 8630.

1. J. Guéna *et al.*, IEEE Trans. on UFFC **57**, 647 (2010).
2. V. Gerginov *et al.*, Metrol. **47**, 65 (2009).
3. K. Szymaniec *et al.*, Metrol. **47**, 363 (2010).
4. F. Levi *et al.*, Metrol. **43**, 545 (2006).
5. S. R. Jefferts *et al.*, IEEE Trans. on UFFC **52**, 2314 (2005).
6. See R. Wynands and S. Weyers, Metrol. **42**, S64 (2005) and references therein.
7. R. Li and K. Gibble, Metrol. **47**, 534 (2010).
8. A. de Marchi *et al.*, IEEE Trans. Instrum. Meas. **37**, 185 (1988).
9. R. Li and K. Gibble, Metrol. **41**, 376(2004).
10. As in [7], b=1 is defined as a $\pi/2$ pulse when averaged over the entire aperture. Since the atomic cloud is small on the upward passage, the atoms experience a larger pulse so our maximum Ramsey fringe contrasts occur at b's slightly smaller than 1,3,5,... .
11. S. Weyers *et al.*, Proc. 21$^{st}$ EFTF and 2007 IEEE Int. Freq. Contr. Symp., 52 (2007).
12. Our cavity's radius is 25 mm, height 26.87 mm, endcap holes 12 mm diameter, and loaded Q 7000. In our model, we fit the cavity height and length of the TM mode filters [1], giving m=0 and 1 modes at 9.1926 and 9.6436 GHz.
13. F. Chapelet *et al.*, Proc. 20th EFTF, (2006).
14. S. Yanagimachi *et al.*, Japan. J. Appl. Phys. **44**, 1468 (2005); M. Kumagai *et al.*, Metrol. **45**, 139 (2008).
15. A small offset was included, due to a problem in the microwave circuitry, which has been corrected.
16. J. Guéna *et al.*, Proc. of the 24$^{th}$ EFTF, (2010).
17. If the sensitivity to perpendicular tilt is significantly different *vs* amplitude (b=1,3,5), it could be used to find zero tilt, but the m=0 DCP shift is a potentially difficult systematic offset
18. We allow the Rb clock to also have an m=1 shift, but 3 times smaller, as measured for tilts along the feeds.
19. We have not evaluated DCP shifts due to machining imperfections in the waveguide cutoffs that could produce large fields and modify the density distribution [7]. These can be eliminated with additional apertures to remove atoms traveling near the cutoff walls.
20. K. Gibble, Phys. Rev. Lett. **97**, 073002 (2006).